\documentclass{article}
\usepackage{ijcai22}

\usepackage{multirow}

\usepackage{times}
\usepackage{soul}
\usepackage{url}
\usepackage[utf8]{inputenc}
\usepackage[small]{caption}
\usepackage{graphicx}
\usepackage{amsmath}
\usepackage{amsthm}
\usepackage{booktabs}
\usepackage{algorithm}
\usepackage{algorithmic}
\usepackage{color}
\definecolor{royalblue}{RGB}{65,105,225} %
\usepackage[pagebackref=true,breaklinks=true,colorlinks,citecolor=royalblue,bookmarks=false]{hyperref}

\title{Hiding Data in Colors: Secure and Lossless Image Steganography \\via Conditional Invertible Neural Networks}

\author{
Yanzhen Ren$^1$
\and
Ting Liu$^1$\and
Liming Zhai$^1$\And
Lina Wang$^1$
\affiliations
$^1$WuHan University
\emails
\{renyz, leeeliu, limingzhai, lnwang\}@whu.edu.cn
}

\begin{document}

\maketitle

\begin{abstract}
{
Deep image steganography is a data hiding technology that conceal data in digital images via deep neural networks.
However, existing deep image steganography methods only consider the visual similarity of container images to host images, and neglect the statistical security (stealthiness) of container images.
Besides, they usually hides data limited to image type and thus relax the constraint of lossless extraction.
In this paper, we address the above issues in a unified manner, and propose deep image steganography that can embed data with arbitrary types into images for secure data hiding and lossless data revealing. First, we formulate the data hiding as an image colorization problem, in which the data is binarized and further mapped into the color information for a gray-scale host image. 
Second, we design a conditional invertible neural network which uses gray-scale image as prior to guide the color generation and perform data hiding in a secure way. Finally, to achieve lossless data revealing, we present a multi-stage training scheme to manage the data loss due to rounding errors between hiding and revealing processes.
Extensive experiments demonstrate that the proposed method can perform secure data hiding by generating realism color images and successfully resisting the detection of steganalysis. Moreover, we can achieve 100\% revealing accuracy in different scenarios, indicating the practical utility of our steganography in the real-world.
}
\end{abstract}

\section{Introduction}
Image steganography is a data hiding technology for concealing the existence of communication, in which the images containing secret data are called container or stego images and the corresponding clean images are called host or cover images.
The image steganography can be generally measured by four criteria: correctness (the revealed data and the secret data are the same), security (the host image and the container image are perceptually and statistically indistinguishable), capacity (the hidden space should be large enough) and robustness (the data revealing should be immune to image processing or noise attacks).
Early image steganography uses hand‐crafted rules  to hide data into host images  \cite{UNIWARD,UED,UERD,MiPOD,CMD,CPP}, requiring abundant domain expert knowledge. Recently, deep learning techniques have been adopted prevalently for image steganography \cite{ddh,ddh_2,Hidden,udh}, which usually consists of a hiding network and a revealing network to perform data hiding and data revealing in an end-to-end pipeline.

The current deep image steganography focuses primarily on the visual quality and capacity of container images, but neglects other important aspects of steganography. First, most deep image steganography only hides data in the form of image, and the other data types, such as text and codes, cannot be directly used for hiding, thus limiting its application in real-world practice. Second, the existing works emphasize more on the visual imperceptibility, and do not consider the statistical security of steganography, which is vulnerable to steganalysis \cite{SRNet,ke}. Third, the deep image steganography cannot guarantee the complete and accurate data revealing, due to the fact that separate hiding and revealing networks are difficult to train to obtain a revealed image identical to a hidden image.


To address the above problems, we propose a new deep image steganography that embeds data with arbitrary types into host images for secure data hiding and lossless data revealing. Different from prior art that produces visually unchanged container images where the hidden images are transferred to high-frequency noises \cite{udh}, we formulate the image steganography as an image colorization problem, for which the gray-scale host images remain unchanged and the hidden data (including images) is generated as color information to form color container images. The rationale behind is that the human visual system is less sensitive to color than luminance, and generating color information can better conceal the steganography traces without changing the texture of the host images.

To ensure that the generated color information is eye-pleasing and also fits the content of gray-scale image, we adopt a conditional invertible neural network (cINN) as a hiding network, which uses the gray-scale image as a condition and transfers the data to colors guided by this condition. Owning to the invertible nature of cINN \cite{NVP,Cinn}, we also use the cINN for revealing, namely the hiding network and the reveal network have the same network architecture and share network parameters. For lossless data revealing, we do not jointly train the hiding network and the reveal network like previous work, and propose a multi-stage training scheme, in which the hiding network is firstly trained alone in initial steps (warm-up stage), and then the hiding network and revealing network are trained in a round manner. 
Our proposed deep image steganography is referred to as Steg-cINN.
We conduct large-scale experiments to demonstrate the effectiveness and advantages of Steg-cINN, which  outperforms the existing works in image quality and statistical security. Moreover, we can achieve 100\% revealing accuracy with the hiding capacity of 2 bpp in spite of rounding errors.
Due to its security and lossless data revealing, it is more flexible and has more extensive applications.

The contributions of this paper are summarized as follows.
\begin{itemize}
\item We propose an image steganography method from the perspective of image colorization based on cINN, which maps the secret data to the color information of container images. The generated color images have high photo-realism for visual security and high undetectability for statistical security.

\item We propose a mapping module that transforms the secret data to latent space required for cINN, enabling data hiding with arbitrary types of data and improving the application scope of our steganography.

\item We propose a multi-stage training scheme that updates the hiding network and revealing network asynchronously, ensuring complete and correct data revealing even under the conditions of rounding errors.
\end{itemize}

\section{Related work}
\label{Related}
\subsection{Image steganography}

Traditional image steganography performs data hiding following a modification principle, for which the host images are slightly modified in a low magnitude, e.g., least significant bit (LSB) modification, to reduce quality degradation and statistical detectability. The typical method is content-adaptive steganography \cite{UNIWARD,UED,UERD,MiPOD,CMD,CPP}, which first assigns a cost to each modified pixel, and then embeds the data while minimizing the sum of costs of all modified pixels using a coding scheme \cite{stc}. The costs and the coding scheme for the traditional image steganography are all designed using man-made rules, requiring sophisticated knowledge and experiences.

With the development of deep learning, researchers begin to use deep neural networks (DNNs) for image steganography. \cite{ddh,ddh_2} propose to directly hide images into images by using three networks: a preparation network for pre-processing the secret image to be hidden, a hiding network for fusing the processed secret image and host image, and a revealing network for revealing secret image. \cite{beforeHidden,Hidden} also use three networks, including a hiding network, a revealing network and a steganalysis work, in which the steganalysis work is used for adversarial training to improve the security. Inspired by universal adversarial examples, \cite{udh} present a universal deep hiding method, which uses a hiding network to transform the secret image to a high-frequency noise image without involving the host image, and the noise image can be superimposed on any host images for data revealing. 
There are also some deep steganography methods based on DNNs \cite{SSGAN,SGAN}, but they use DNNs to synthesize proper host images, and the hiding and revealing processes are similar to traditional image steganography.

The above image steganography hide the secret data into all channels of host images (color or gray-scale), while our method transfer the secret data to color channels for a gray-scale host image. The previous deep image steganography consists of multiple networks based on convolutional neural networks. As a contrast, our method only use one network (hiding network and revealing network share the same architecture) based on a conditional invertible neural network.

\subsection{Conditional invertible neural network(cINN)}\label{cINN}

Invertible neural network (INN) is a neural network that has a bijective and invertible construction \cite{NICE,NVP,Glow}.
The INN models inverse problems within a single network and its invertible architecture enables efficient bidirectional training.
As an important variant of INN, the conditional INN (cINN) extends the applications of INN, such as guided image generation \cite{Cinn} and audio generation conditioned on mel-spectrogram \cite{WaveGlow}.
In this paper, we take the advantage of cINN’s bijective  and invertible properties for designing our steganography.

There have been some works using INNs for steganography, including hiding images in images \cite{ISN}, hiding video in audio \cite{Hivia} and hiding binary data through audio generation \cite{Steg-TTS}. These INN based steganography can correctly reveal the secret data due to the invertible structure. However, they can not ensure accurate revealing  when container images (or audio) suffer from rounding operations during multimedia storage. By contrast, we propose a multi-stage training scheme to achieve accurate data revealing in spite of rounding errors.

\section{Methodology}
\begin{figure*}[ht]
    \centering
    \includegraphics[scale=0.35]{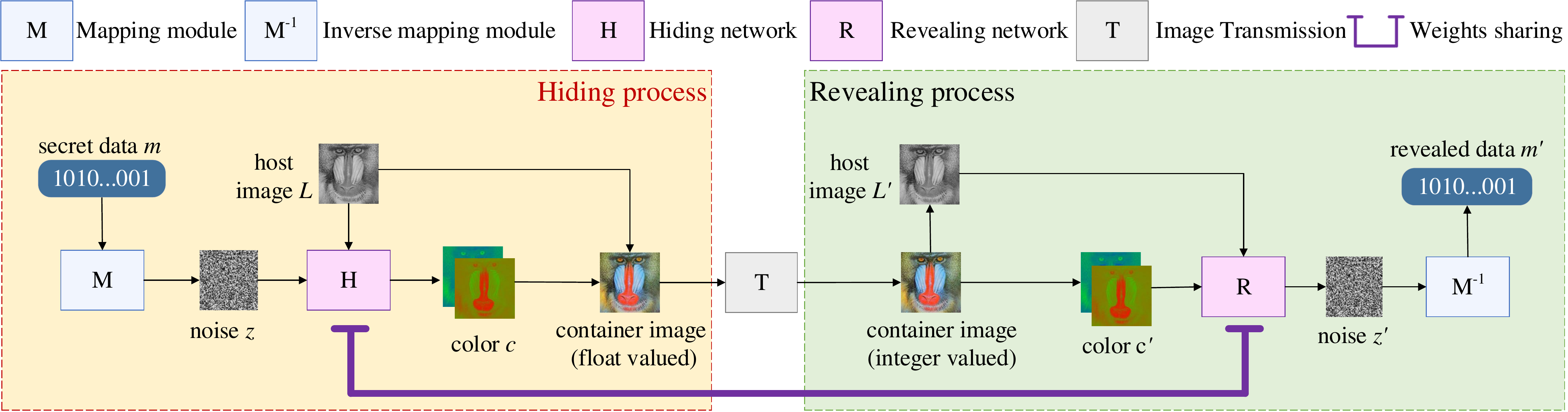}
    \caption{Framework of our deep image steganography Steg-cINN. The framework consists of a Hiding process and a Revealing process. Note that M and M$^{-1}$ are shown in algorithm\ref{M_hide}-\ref{M_extract}. Hiding network and revealing network are both based on cINN as shown in Fig.\ref{fig:cINN_cACL}.
    }
    \label{fig:steg_Cinn_framework}
\end{figure*}
\begin{figure}[ht]
    \centering
    \includegraphics[scale=0.3]{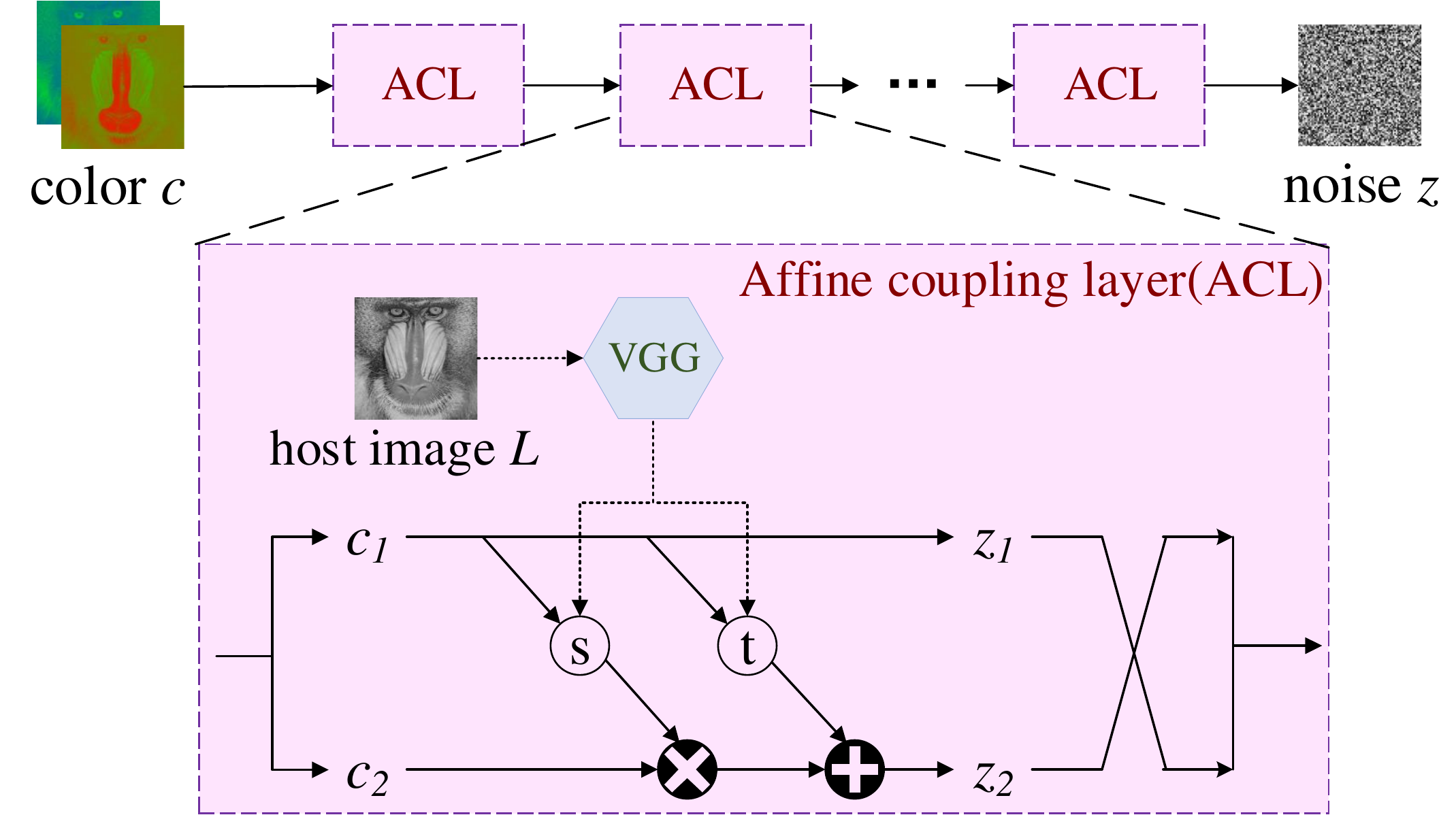}
    \caption{Architecture of conditional invertible neural network(cINN). This network architecture is used for both hiding network and revealing network.}
    \label{fig:cINN_cACL}
\end{figure}

\subsection{Framework of Steg-cINN}

The framework of the proposed Steg-cINN is shown in Fig.\ref{fig:steg_Cinn_framework}, including a hiding process and a revealing process.
For the hiding process, the hiding network takes the gray-scale image as a conditional guidance, hides the binarized secret data during the coloring process, and synthesizes the color container image. For the revealing process, the revealing network first decomposes the container image to obtain the gray-scale image and color information, and then reveals the binarized secret data from the color information.

\paragraph{Hiding process.}

For any types of secret data, it should be binarized to bit-stream before hiding. Then the mapping module $M$ maps the binarized secret data $\mathbf{m}$ into latent variable $z$ which follows a standard normal distribution.
For a color host image, it will be converted to gray-scale image using a Lab color space.
The hiding network $H$ transfers the latent variable $z$ into new color information $c$ under the guidance of gray-scale image $L$, where $L$ is pre-processed by a pre-trained VGG \cite{VGG} to make the color information better match the semantic information of the gray-scale image.
The structure of hiding network $H$ is based on cINN (shown in Fig.\ref{fig:cINN_cACL}), which is made up of multiple affine coupling layers (ACL).
In each ACL, $s$ and $t$ are arbitrary DNNs.
For $L$ and $c$, they are concatenated and converted to RGB image through color space conversion. After image storage with rounding operations, the integer valued image will be sent as a container image.

\paragraph{Revealing process.}
The revealing process is the inverse of hiding process.
The container image is first converted from RGB to Lab color space to obtain $L'$ and $c'$, where $L'$ represent reconstructed gray-scale image and $c'$ represent reconstructed color information.
Then the revealing network $R$ transfers $c'$ to latent variable $z'$ under the guidance of $L'$. The revealing network $R$ is also based on cINN, and it has the same architecture and shares weights with the hiding network $H$.
Finally, the inverse mapping module $M^{-1}$ maps $z'$ to binary bit-stream $m'$, which can be further converted to its original data type.

\begin{figure}[ht]
    \centering
    \includegraphics[scale=0.36]{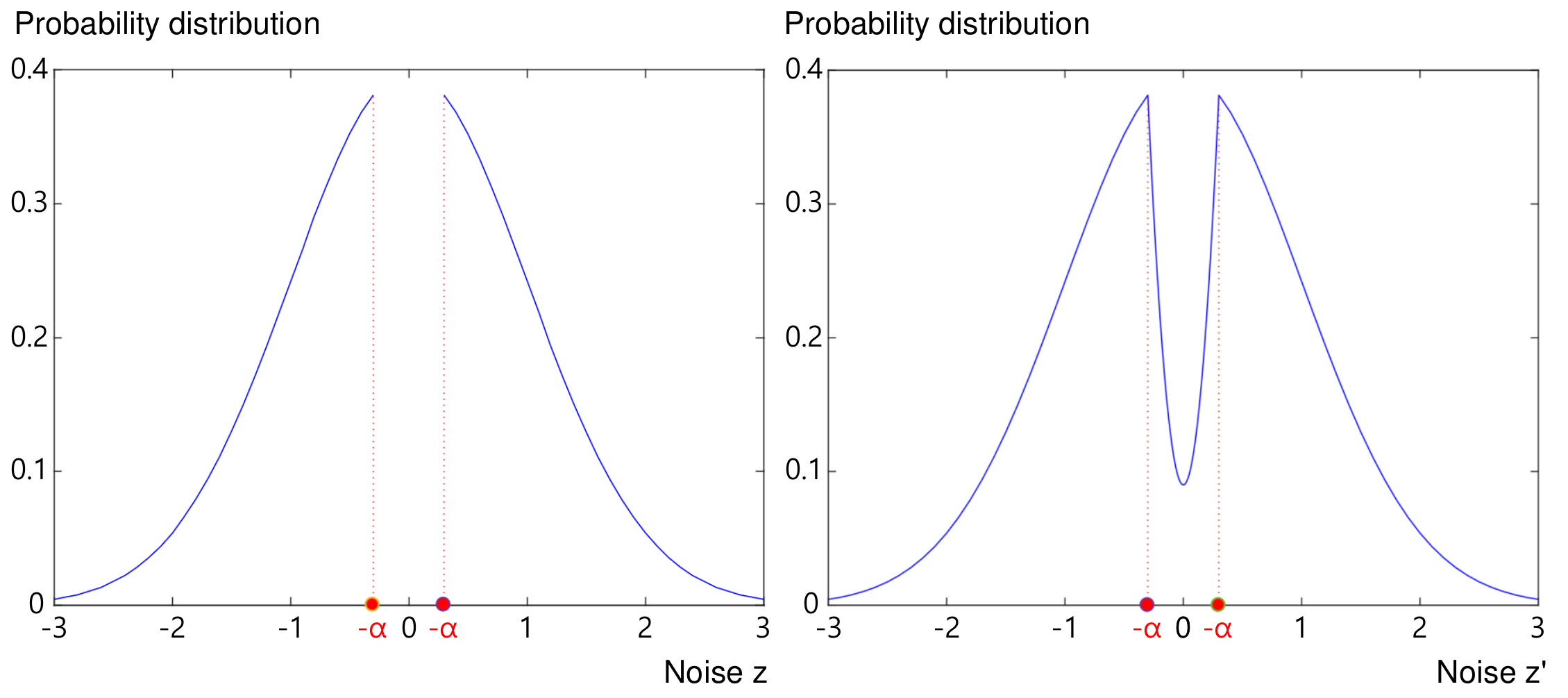}
    \caption{Two kinds of probability distribution. The left is the distribution of noise $z$ in hiding process; and the right is the distribution of noise $z'$ in revealing process.}
    \label{fig:ZandZ_hat}
\end{figure}

\paragraph{Mapping module and inverse mapping module.}
The mapping module $M$ and inverse mapping module $M^{-1}$ play important roles in hiding process and revealing process.
The main idea is to let the sign of $z$ represent $\mathbf{m}$, which is a binary bit stream that follows a uniform distribution, and $z$ is a latent variable that follows a standard normal distribution $\mathcal{N}(0,1)$.
For error tolerance, an interval parameter $\alpha$ is defined so that z $\in$ [$-\alpha$,$\alpha$] is rejected when sampling $z$. The $\alpha$ is set to 0.1 by default.
Therefore, in the hiding process, there is a ``gap" of width $2\alpha$ in the distribution of $z$, as shown in the left figure of Fig \ref{fig:ZandZ_hat}. 
Considering that the rounding operations in image storage will cause information loss, in the revealing process, the extracted $z'$ may be as shown in the right figure of Fig \ref{fig:ZandZ_hat}. In this way, as long as the signs of $z$ and $z'$ are the same, the $\mathbf{m}$ can be accurately extracted.
The pseudo-code of $M$ and $M^{-1}$ are shown in Algorithm \ref{M_hide} and \ref{M_extract}, respectively.

\begin{algorithm}[tb]
\caption{Mapping module $M$}
\label{M_hide}
\textbf{Input}: Binary data $\mathbf{m}=\{0,1,...,0,1\}$, $\alpha = 0.1$    \\
\textbf{Output}: Latent variable $z$
\begin{algorithmic}[1]
\FORALL {$b$ in $\mathbf{m}$}
	\IF {$b==0$}
		\STATE Sample $z$ from $\mathcal{N}(0,1)$ until $z < -\alpha$
	\ENDIF
	\IF {$b==1$}
		\STATE Sample $z$ from $\mathcal{N}(0,1)$ until $z > \alpha$
	\ENDIF
\ENDFOR
\end{algorithmic}
\end{algorithm}

\begin{algorithm}[tb]
\caption{Inverse mapping module $M^{-1}$}
\label{M_extract}
\textbf{Input}:  Latent variable $z'$ \\
\textbf{Output}: Revealed binary data $\mathbf{m'}$
\begin{algorithmic}[1]
\STATE Initialize $\mathbf{m}'$
\FORALL{noise $n'$ in $z'$}
\IF {$n' < 0$}
\STATE Extract bit 0 into $\mathbf{m}'$
\ELSE
\STATE Extract bit 1 into $\mathbf{m}'$
\ENDIF
\ENDFOR
\end{algorithmic}
\end{algorithm}

\subsection{Training of Steg-cINN}

The training of Steg-cINN is divided into two stages.
For the first stage, the target is to train a generative model which generates colors that look natural and real. Similar to the normal cINN \cite{Cinn}, the training strategy is maximum likelihood training.
For the second stage, the Steg-cINN is trained in a round-based manner to enhance its revealing capability under a rounding error condition.

\paragraph{Training in the first stage.}

Let the Steg-cINN be denoted by $f(c;L)$, the learning objective is to make $z=f(c;L)$ close to $\pi(z)$, where $c$ is color information decomposed from RGB images using a Lab color space model \cite{Lab}, $L$ is a gray-scale image, and $\pi(z)$ is a standard normal distribution.
Since the Steg-cINN is invertible, as long as $z=f(c;L)$ is close to $\pi(z)$, $c=f^{-1}(z;L)$ will be close to $q(c)$, where $q(c)$ is the distribution of real-world image colors $c$ (relevant to specific datasets). According to the idea of maximum likelihood estimation (MLE), the loss function in the first stage is designed as a negative log-likelihood (NLL) loss
\begin{equation}\label{stageI}
\begin{aligned}
 {\mathcal{L}}_1 = -\log q(c;L) = -\log \left(\pi(z) \left| det \frac{\partial z}{\partial c} \right|\right)
\end{aligned}
\end{equation}
 where $\left| det \frac{\partial z}{\partial c} \right| $ is the determinant of Jacobian matrix.

\paragraph{Training in the second stage.}

\begin{algorithm}[tb]
\caption{Round-based training}
\label{stageII-pseudocode}
\textbf{Input}: Steg-cINN model trained in the first stage, binary data $\mathbf{m}=\{0,1,...,0,1\}$, gray-scale image $L$ \\
\mbox{\textbf{Output}: Hiding network $H$ and revealing network $R$}
\vspace{-12pt}
\begin{algorithmic}[1]
\STATE Initialize the weights of the hiding network $H$ and revealing network $R$ with the Steg-cINN model
\FOR{$r$ in rounds}
    \STATE The hiding network $H$ generate float valued container images from $\mathbf{m}$ and $L$
    \STATE Save the float valued container images to obtain integer valued RGB images
    \FOR{$i$ in iterations}
        \STATE Train the revealing network $R$ and update its weights
    \ENDFOR
    \STATE Copy the weights of $R$ to $H$
\ENDFOR
\end{algorithmic}
\end{algorithm}

The training in the first stage empowers the Steg-cINN with generative ability for synthesizing colors, but the Steg-cINN still lacks of hiding and revealing ability. The previous DNN- or INN-based steganography directly jointly or bidirectionally trains the hiding network and revealing network, but they neglect the rounding error existed in the image saving operation, which cause information loss and thus may further cause inaccurate revealing. one possible solution is to round the images during training, but it will affect the the backward propagation of training.

To solve this problem, we propose a round-based training strategy. The training consists of multiple rounds. In each round, the weights of hiding network are frozen, and the hiding network only do the inference process. Specifically, the mapping module samples $z$ from $\mathcal{N}(0,1)$, and the $z$ is fed to the hiding network to generate float valued container images. After rounding operation, the integer valued images are fed to the revealing network and then the backward propagation is performed to update the weights of the revealing network. The training of revealing network is iterated until a desired number of iterations is reached. At the end of each round, the weights of the revealing network will be copied to the hiding network. The round-based training process is detailed in Algorithm \ref{stageII-pseudocode}.

The loss function in the second stage is defined as
\begin{equation}\label{stageII}
\begin{aligned}
 {\mathcal{L}}_2 &= -\log q(c';L') + {\left\| z- z' \right\|_2} \\
    &= -\log \left(\pi(z') \left| det \frac{\partial z'}{\partial c'} \right|\right) + {\left\| z- z' \right\|_2}
\end{aligned}
\end{equation}
where the first term is the NLL loss similar to that in the first stage, and the second term is a reconstruction loss to ensure the revealing accuracy. The $c'$ and $L'$ are the revealed color information and recovered gray-scale image.

\section{Experiments}
\label{exp}
\subsection{Experimental Setup}
\paragraph{Datasets.}
We use two datasets COCO \cite{Coco} and BossBase \cite{Boss} to evaluate the performance of our Steg-cINN. 
The COCO contains 80 object categories, and we randomly select 10,000 images for experiments.
The BOSSBase consists of 10,000 uncompressed images coming from seven different cameras.
To test the transferbility of deep models, the training is performed on COCO and the testing is performed on BossBase.
Similar to previous works \cite{ddh,udh,Hidden}, all the images are resized to the resolution of 128×128.

\paragraph{Baselines.}
We adopt three deep steganography methods, i.e., DDH \cite{ddh}, UDH \cite{udh} and HiDDeN \cite{Hidden}, for comparison. The DDH and UDH hide images into images, while the HiDDeN hide bit-streams into images. All the baselines are implemented using their default settings.

\paragraph{Metrics.}
The performance of deep steganography is evaluated by four metrics:
visual quality, revealing accuracy, hiding capacity and statistical security. Since the container images of Steg-cINN are generated unsupervisely, we use no-reference image quality assessment Brisque \cite{Brisque} and hyperIQA \cite{hyperIQA} to measure the visual quality. The revealing accuracy refers to the ratio of amount of correctly recovered bits to the total amount of hidden bits. The hiding capacity is expressed in bpp (bit per pixel), which is the maximum amount of the secret bits that can be hidden in the host image.
The statistical security is evaluated by three steganalysis models XuNet\cite{Xu}, YedroudjNet\cite{Ye} and KeNet\cite{ke}, for which lower detection accuracy means high statistical security.

\paragraph{Implementation Details.}
Our Steg-cINN is implemented with PyTorch, and an Nvidia Tesla V100 GPU is used for acceleration.
We selected Adam as optimizer with $\beta_1$ = 0.9 and $\beta_2$ = 0.999.
For the round-based training, the round number and the iteration number in each round is 5 and 4000.
In each round, the learning rate starts from 0.0001 and is divided by 5 when the training loss plateaus. The batch size is set as 48.
Our network contains 30 affine coupling layers, each of them uses convolutional neural networks as the $s$, $t$ transformation (see $s$ and $t$ in Fig. \ref{fig:cINN_cACL}), respectively. The details of the network architecture are provided in supplementary material.

\subsection{Qualitative Evaluation of Visual Quality}\label{exp1}

\begin{figure*}[t]
    \centering
    \includegraphics[width=1.49\columnwidth]{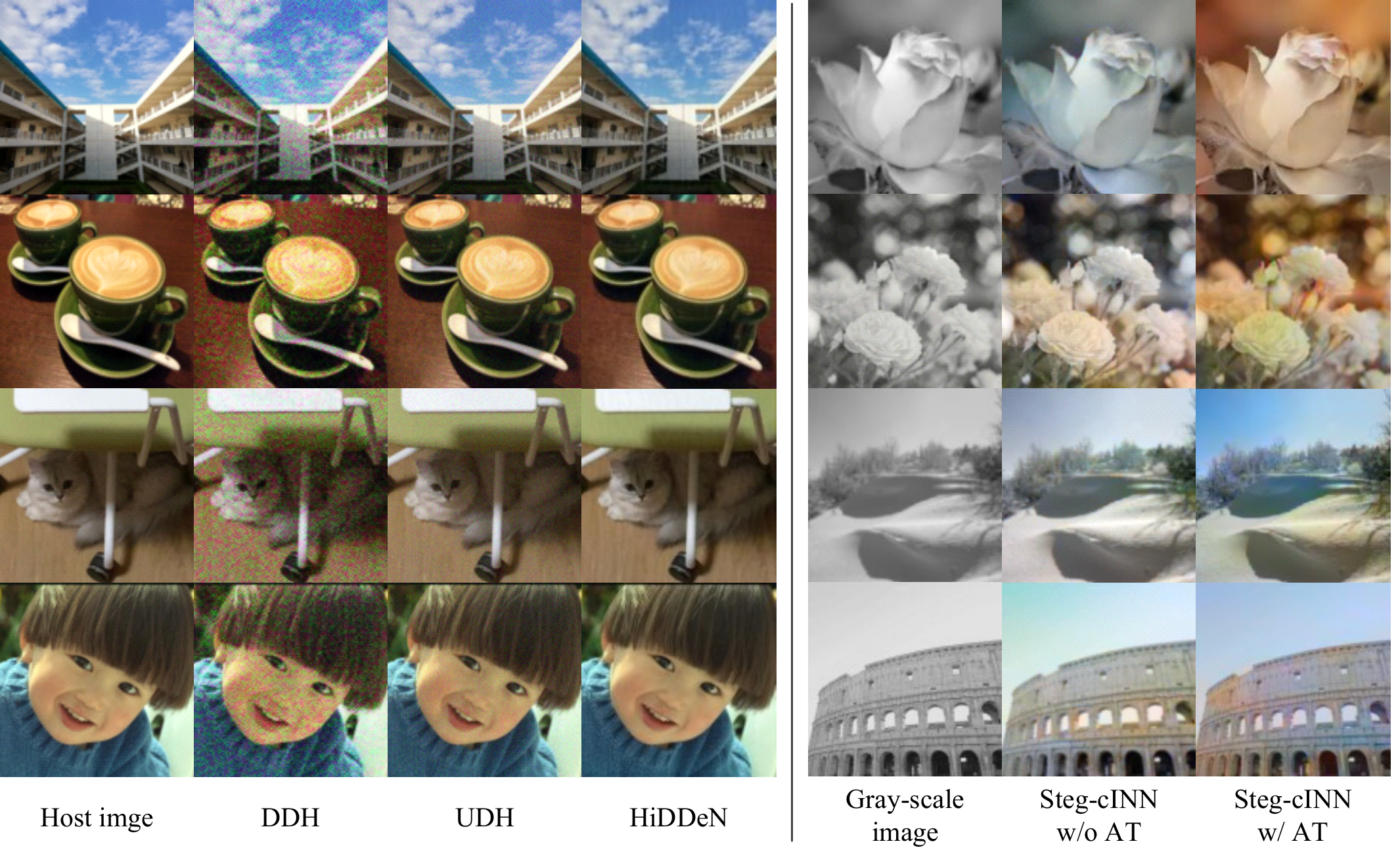}
    \caption{Qualitative comparison of visual quality.}
    \label{fig:experiment-visual}
\end{figure*}

Fig.\ref{fig:experiment-visual} shows the visualization of host images and container images for baselines and our Steg-cINN.
The hiding capacity of DDH, UDH and HiDDeN are 1.0, 1.0 and 0.0018 bpp, and the hiding capacities of the two versions of Steg-cINN are both 2.0 bpp.

We can see from the left figure in Fig. \ref{fig:experiment-visual} that the container images of DDH and UDH have some visual distortions due to their large hiding capacity. The container images of HiDDeN are very similar to the host images, since the hiding capacity is very low.
We observe from the right figure in Fig. \ref{fig:experiment-visual} that compared with the gray-scale images, the Steg-cINN without round-based training (Steg-cINN w/o RT) generates light-colored images, but the Steg-cINN with round-based training (Steg-cINN w/ RT) can produce more colorful images, which better match the semantic content of gray-scale images, demonstrating the effectiveness of round-based training in photo-realism.

\subsection{Quantitative Evaluation of Visual Quality}\label{exp2}

\begin{table}[]
\centering
\begin{tabular}{lcc}
\toprule
Deep Steganography & Brisque $\downarrow$ & HyperIQA $\uparrow$ \\ 
\midrule
DDH               & 49.97  & 20.27 \\
UDH               & 18.35  & 36.11 \\
HiDDeN            & 16.38  & 33.71 \\
Steg-cINN w/o RT & 20.78  & 31.44 \\
Steg-cINN w/ RT   & 19.74  & 38.46 \\ \bottomrule
\end{tabular}
\caption{Quantitative comparison of visual quality.}
\label{tab:comparison-NR-IQA}
\end{table}

We report the scores of Brisque \cite{Brisque} and hyperIQA \cite{hyperIQA} for different deep steganography in Table~\ref{tab:comparison-NR-IQA}.
We observe that the Steg-cINN with round-based training (Steg-cINN w/ RT) achieves comparable or better results than existing methods in terms of Brisque, and performs the best overall in terms of hyperIQA. 
The visual performance of Steg-cINN without round-based training (Steg-cINN w/o RT) is inferior to Steg-cINN w/ RT, which is consistent with that in Section \ref{exp1}.
It should be noted that the baselines are all supervised learning, while our Steg-cINN is based on a unsupervised scheme. Besides, the Steg-cINN also uses a higher hiding capacity for visual comparison.

\subsection{Revealing Accuracy}\label{exp3}

\begin{table*}[t]
\footnotesize
  \newcommand{\tabincell}[2]{\begin{tabular}{@{}#1@{}}#2\end{tabular}}
  \centering
  {
  \begin{tabular}{lcccccc}
  \toprule
  \multirow{2}{*}[-2pt]{\tabincell{c}{Deep\\Steganography}}
  & \multirow{2}{*}[-2pt]{\tabincell{c}{Capacity $\uparrow$\\(bpp)}}
  & \multicolumn{2}{c}{Revealing Accuracy $\uparrow$}
  & \multicolumn{3}{c}{Detection Accuracy $\downarrow$} \\
  \cmidrule(l){3-4} \cmidrule(l){5-7}
  &  &  w/o rounding error & w/ rounding error & XuNet & YedroudjNet & KeNet \\
  \midrule
  \multirow{2}{*}{DDH}
  & 1.00 &  98.21 & 69.23 & 100.00 & 100.00 & 100.00 \\
  & 0.25 & 100.00 & 68.26 &  97.37 &  99.56 & 100.00 \\
  \midrule
  \multirow{2}{*}{UDH}
  & 1.00 &  89.95 & 75.96 & 98.31 & 97.62 & 99.04 \\
  & 0.25 & 100.00 & 67.30 & 97.56 & 96.37 & 98.80 \\
  \midrule
  HiDDeN & 0.0018 & 100.00 & 82.69 & 95.18 & 93.12 & 96.88 \\
  \midrule
  Steg-cINN w/o RT
  & 2.00 & 100.00 & 89.42 & 50.34 & 49.59 & 56.28 \\
  Steg-cINN w/ RT  & 2.00 & 100.00 & 100.00 & 52.68 & 52.18 & 56.91 \\
  \bottomrule
  \end{tabular}
  }
  \caption{The capacity, revealing accuracy (in \%) and detection accuracy (in \%) of different deep steganography.}
  \label{tab:acc}
\end{table*}


There are two scenarios for data revealing, one is ideal scenario where the container images are stored and transmitted without any information loss when rounding operation is not considered, and the other one is the practical scenario where the rounding error of container images occurs. The revealing accuracy for the two scenarios are reported in the column 3--4 of Table~\ref{tab:acc}.

We observe that for the ideal scenario, the previous methods can obtain 100\% revealing accuracy only at low hiding capacities, and their revealing accuracy decreases with increasing the hiding capacities. As a contrast, our two versions of Steg-cINN can achieve 100\% revealing accuracy even at higher hiding capacities. For the practical scenario, all the deep steganography will suffer from the information loss problem, so we use BCH error correction code to enhance the revealing accuracy. We can see that even with the aid of error correction code, the revealing accuracy of previous methods still decreases sharply. The Steg-cINN w/o RT achieves higher revealing accuracy than previous methods, and the Steg-cINN w/ RT can gain 100\% revealing accuracy, demonstrating the high revealing ability of our method.

\subsection{Statistical Security}\label{exp4}


We report the steganalysis results for different deep steganography in the last three columns of Table~\ref{tab:acc}. We can see that the detection accuracy for DDH, UDH and HiDDeN all exceed 93\%, indicating that the steganalysis can accurately distinguish between host images and container images and the existing deep steganography lacks of safety in a statistical sense.
By contrast, the detection accuracy for our Steg-cINN all around 50\%, meaning that the Steg-cINN can obtain perfect statistical security.
The reason behind is that the Steg-cINN pursues the statistical similarity of host images and container images.
While DDH, UDH and HiDDeN target at the visual similarity of host images and container images, so they are easy to be detected.



\subsection{Ablation Study}

\begin{table}[t]
\footnotesize
\centering
{
\begin{tabular}{lcc}
\toprule
\multirow{2}{*}{\begin{tabular}[c]{@{}c@{}}Deep\\ Steganography\end{tabular}} & \multicolumn{2}{c}{Revealing Accuracy $\uparrow$} \\ \cmidrule(l){2-3} 
 & w/o rounding error & w/ rounding error \\ 
\midrule
Steg-cINN w/o RT  & 100.00 & 89.42  \\
Steg-cINN (round=1) & 100.00 & 93.26 \\
Steg-cINN (round=2) & 100.00 & 95.19 \\
Steg-cINN (round=3) & 100.00 & 98.07 \\
Steg-cINN (round=4) & 100.00 & 99.03 \\
Steg-cINN (round=5) & 100.00 & 100.00 \\
\bottomrule
\end{tabular}
}
\caption{The revealing accuracy (in \%) of the proposed Steg-cINN with different round numbers.}
\label{tab:ablation}
\end{table}

The round-based training is essential for data revealing in a rounding error scenario, and the round number determines the revealing accuracy. To investigate how the round number affects the data revealing, we perform an ablation experiment to select the proper round numbers. The revealing accuracy of Steg-cINN with different rounds is provided in Table \ref{tab:ablation}.

We can see from Table \ref{tab:ablation} that the round number does not affect the revealing accuracy in a ideal scenario. For the practical scenario where the rounding error exists, the revealing accuracy increases with the increase of round number. The revealing accuracy reaches 100\% when the round number increase to 5, so we set the round number as 5 for Steg-cINN w/ RT in our previous experiments. Note that even using one round can significantly enhance the Steg-cINN's revealing accuracy (from 89.42\% to 93.26\%, compared with Steg-cINN w/o RT), demonstrating the effectiveness of our proposed training strategy. 

\section{Conclusion}
This paper proposes Steg-cINN, a deep image steganography that can embed data with arbitrary types into images for secure data hiding and lossless data revealing. 
We regard the image steganography as a color colorization process, in which the hidden secret data is disguised by the normal color generation.
For this purpose, we design a conditional invertible neural network for deep image steganography, which hides data guided by gray-scale images.
Meanwhile, the Steg-cINN is enhanced by a multi-stage training scheme, where the hiding network and revealing network are trained in a round manner, which ensures accurate data revealing in spite of rounding errors.
The experiments show that the proposed method achieves competitive results with existing methods in visual quality by qualitative and quantitative comparisons.
Furthermore, the proposed method gains higher revealing accuracy and hiding capacity than existing methods.
Last but not least, the proposed method can also guarantee statistical security, outperforming existing methods.
Our future work will focus on improving the robustness of the data hiding, such as resisting heavy image compression and various image processing operations.


\end{document}